\documentclass[12pt]{article}
\usepackage[margin=1in]{geometry}
\usepackage{amsfonts,amsmath,amsthm,array,amssymb}
\usepackage[titletoc,title]{appendix}
\usepackage{hyperref}
\usepackage[square,sort,comma,numbers]{natbib} 
\usepackage{graphicx}
\usepackage{subfigure}
\usepackage{longtable}
\usepackage[margin=1in]{geometry}
\usepackage{setspace}
\usepackage{float}

\usepackage{color}

\date{}

\begin{document}

\setstretch{1}
\title{Modeling Eating Behaviors: the Role of Environment and Positive Food Association Learning via a \textit{Ratatouille} Effect}

\author{Anarina L. Murillo$^1$, Muntaser Safan$^{1, 2, 3}$, Carlos Castillo-Chavez$^1$, \\ Elizabeth D. Capaldi-Phillips$^4$, Devina Wadhera$^4$ \\ 
\footnotesize  $^{1}$Simon A Levin Mathematical, Computational and Modeling Sciences Center,\\
\footnotesize Arizona State University, Tempe, AZ, USA\\
\footnotesize $^{2}$ Mathematics Department, Faculty of Science, Mansoura University, Mansoura, Egypt\\
\footnotesize $^{3}$ Department of Mathematical Sciences, Faculty of Applied Science, Umm Al-Qura University, \\
\footnotesize 21955 Makkah, Saudi Arabia\\
\footnotesize $^{4}$ Conditioned Feeding Lab, Behavioral Neuroscience, Department of Psychology, \\
\footnotesize Arizona State University, Tempe, AZ, USA   \\
\footnotesize Emails: Anarina.Murillo@asu.edu, \footnotesize Muntaser.Safan@asu.edu,  \footnotesize ccchavez@asu.edu, \\ \footnotesize  Betty.Phillips@asu.edu, \footnotesize Devina.Bajaj@asu.edu}

\maketitle 

\begin{abstract}
Eating behaviors among a large population of children are studied as a dynamic process driven by nonlinear interactions in the sociocultural school environment. The impact of food association learning on diet dynamics, inspired by a pilot study conducted among Arizona children in Pre-Kindergarten to 8th grades, is used to build simple population-level learning models. 
Qualitatively, mathematical studies are used to highlight the possible ramifications of instruction, learning in nutrition, and health at the community level. Model results suggest that nutrition education programs at the population-level have minimal impact on improving eating behaviors, findings that agree with prior field studies. Hence, the incorporation of food association learning may be a better strategy for creating resilient communities of healthy and non-healthy eaters. A \textit{Ratatouille} effect can be observed when food association learners become food preference learners, a potential sustainable behavioral change, which in turn, may impact the overall distribution of healthy eaters. In short, this work evaluates the effectiveness of population-level intervention strategies and the importance of institutionalizing nutrition programs that factor in economical, social, cultural, and environmental elements that mesh well with the norms and values in the community. 
\end{abstract}


 \section{Introduction}


The prevalence of childhood obesity has doubled among 2-to-5-year-olds (5-7\% to 10.4\%) and tripled for both 6-to-11-year-olds (6.5\% to 19.6\%) and 12-to-19-year-olds (5\% to 18.1\%) from $1971-1974$ to $2007-2008$ \citep{Ogden(2010a)}. Childhood obesity can increase risk of cardiovascular disease \citep{ammerman2002efficacy,boeing2012critical,ness1997fruit} and cancer \citep{ammerman2002efficacy,block1992fruit,lipkin1999dietary}, two leading causes of premature mortality and physical morbidity in adulthood \citep{reilly2011long}. Many national efforts, such as the United States Department of Agriculture's (USDA) implementation of the ``My Plate" guidelines \citep{USDAmyplate} in schools, aim to alter the eating dynamics of young individuals \citep{ammerman2002efficacy}. These state-mandated guidelines impact the diets of those who eat lunch (60\%) and breakfast (37\%) at their schools \citep{USDAFNS2005}, or  99\% and 78\% of public schools who participate in the National School Lunch and Breakfast Programs, respectively \citep{Fox(2004),kaphingst2006role}. In short, children, in the early stages of developing their eating habits, consume most of their daily food (19 to 50\% or more) in schools \citep{Gleason(2001),kaphingst2006role} and are members of a captive audience (10 years, 9 months, and 5 days per week) \citep{ADA(1999),perez2001school}. Hence, a better understanding of the overall effectiveness of these programs and the access to a captive audience is necessary for improving the overall health of children.
 
\begin{figure}[H]
\centering
\includegraphics[scale=.256]{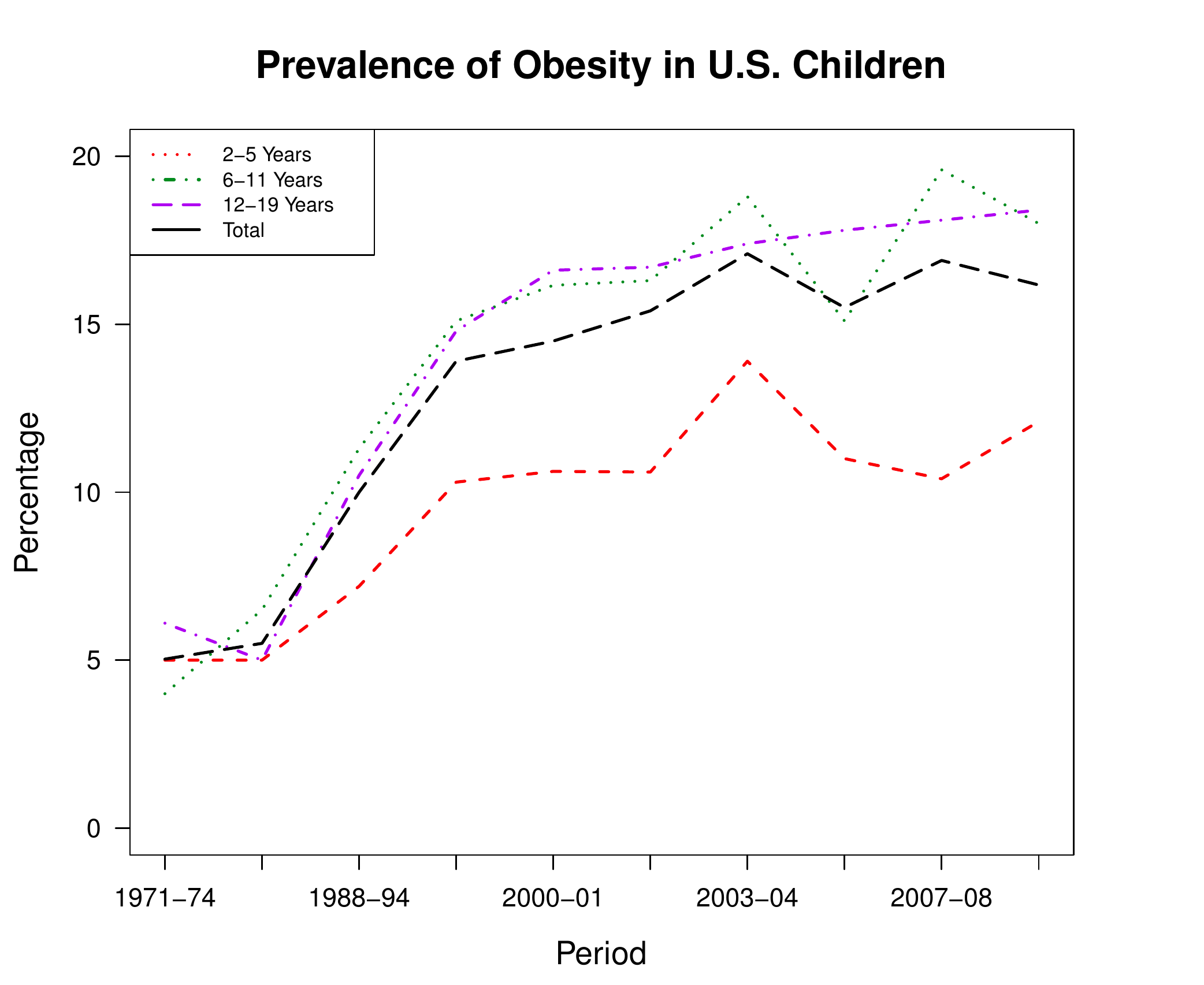} 
\hspace{0.1px}
\includegraphics[scale=.265]{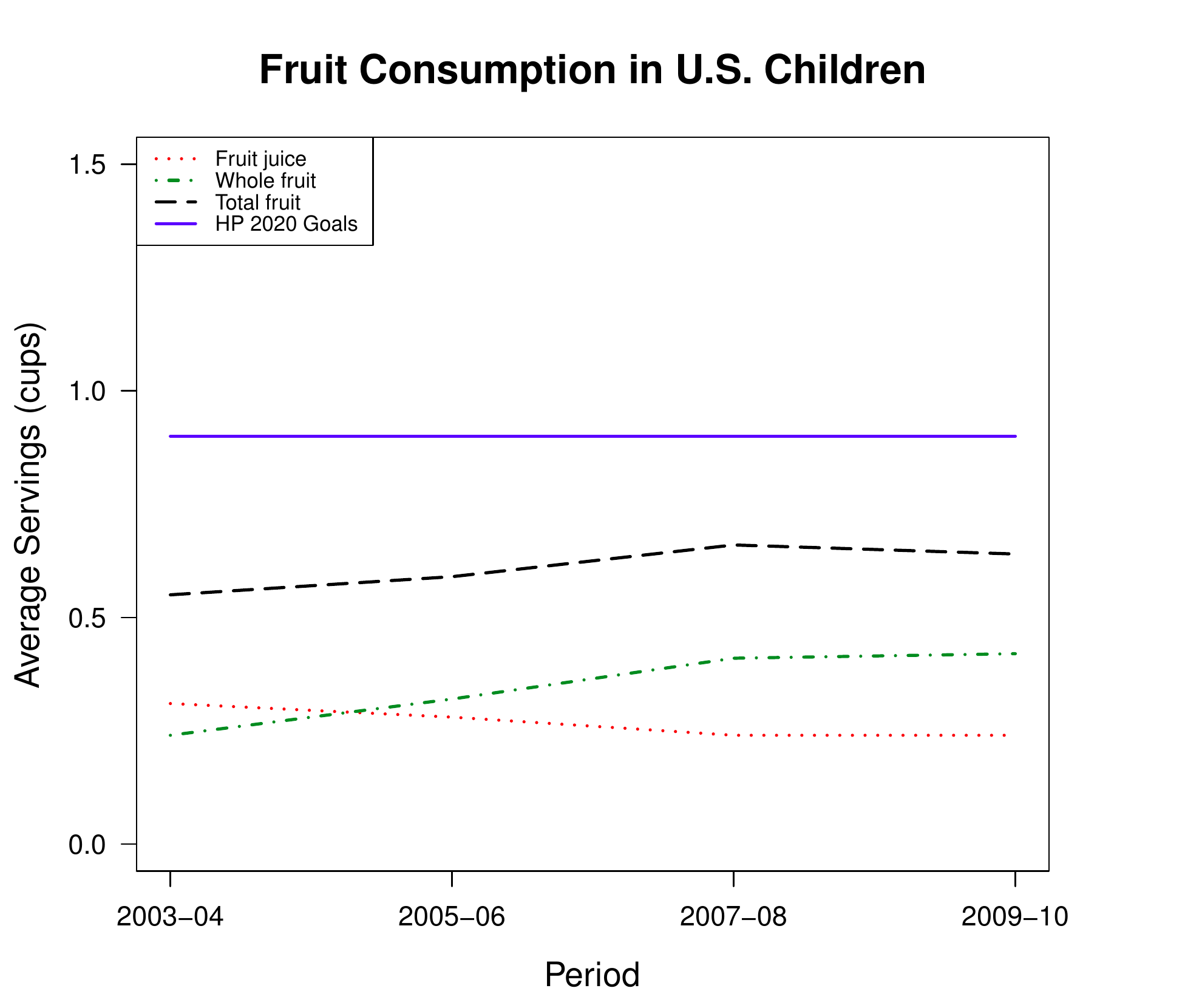} 
\hspace{0.1px}
\includegraphics[scale=.265]{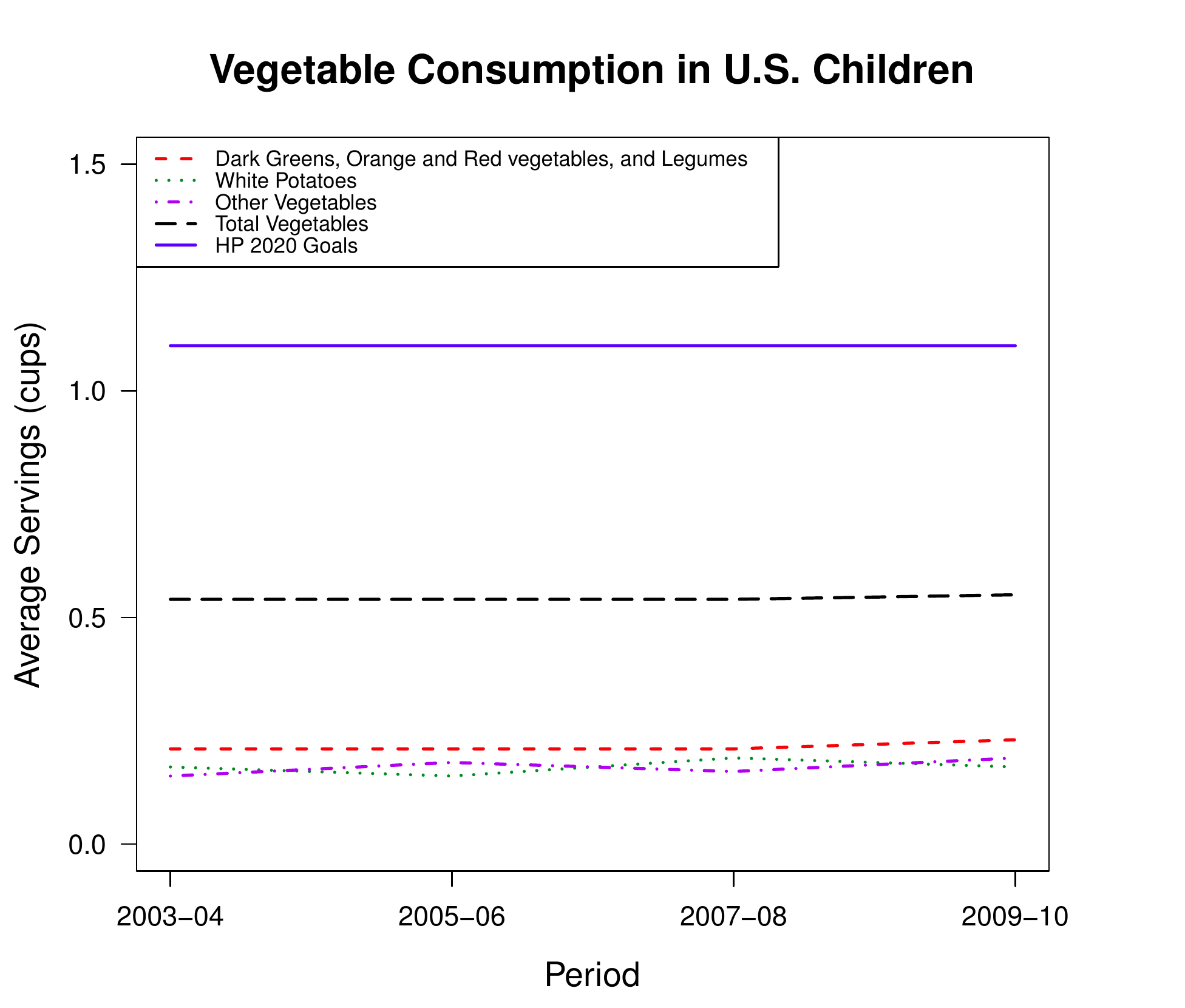}
\caption{Prevalence of childhood obesity (left). Fruit (middle) and vegetable (right) consumption in U.S. children \citep{CDC(2014),Ogden(2010a)}.}
\label{fig:fruitvegconsumption}
\end{figure}

In this paper, we aim at shedding some light on the connections between key identified factors \citep{ammerman2002efficacy,blanchette2005determinants,katz2009school,lytle1995changing} that shape eating behaviors at the population-level via contagion mathematical models, within a social-ecological framework \citep{mcleroy1988ecological}. Although schools are ideal for institutionalizing nutrition programs, a huge step in the fight against obesity-related illness, childhood obesity is still an issue and consumption of fruits and vegetables among children is low (see Figure~\ref{fig:fruitvegconsumption}). Using the film \textit{Ratatouille} as a metaphor and the study by \citeauthor*{wadhera2015perceived} \citeyearpar{wadhera2015perceived}, we investigate the significance of the ``\textit{Ratatouille}" effect, that is the impact of recreating `positive' childhood eating experiences, memories, and their connection with the process of building healthy eating habits. Food preference learning has been identified as a possible influential method for developing healthier eating habits by modifying taste, the strongest predictor of children's food consumption \citep{birch1979preschool,domel1993measuring,perry2004randomized,story2002individual}. Although well-studied in experimental settings, its impact is not well-understood at the population-level, and hence, we investigate this phenomenon on the diet dynamics of young individuals in this work.

 \singlespacing
 
\section{Eating Behaviors in School Settings}
The study of the diet dynamics of individuals at the population-level have been rarely addressed in the literature (but see \cite{evangelista2004usa,frerichs2013modeling,gonzalez2010age1,jodar2008modeling}). Building a population-level model from the knowledge that we have gathered on the daily decisions of individuals is rather challenging just as it is the construction of an epidemiological model from the study of an individuals immunological (level of the cell) response to a disease invasion. Our eating behaviors, that is, why we eat certain foods, how much to eat, when to eat, and how to eat these foods, are governed by biological, sociocultural, and psychosocial factors that are learned in a variety of settings. In this work, we assume that there are three population-level components involved on the diet-dynamics of individuals within a community. The first involves the impact of dietary programs (health awareness, communication, and skill-building) that tend to be temporary and often associated with high levels of recidivism \citep{cullen1998measuring,domel1993measuring,pyle2006fighting}. The second would be the social environment, here modeled simply via the day-to-day interactions among individuals with different diets. The unpalatability of healthy foods make their systematic consumption difficult, however, social and behavior-based elements have been shown effective; such as, hands-on curriculum activities (classroom lessons, taste-testing, cooking lessons), parental involvement, school gardening, peer modeling, or rewards \citep{ammerman2002efficacy,birnbaum2002differences,lowe2004effects,lytle1995changing,perry2004randomized,story2002individual}. The third includes the physical environment, here availability and accessibility of healthier foods changes due to the nutrition programs implemented in the schools \citep{cullen2003availability,perez2003nutrition,story2008creating,van2007systematic}. Despite our understanding of these factors, the efficacy of these interventions vary and so, more work is needed in order to fully assess their impact on the diet dynamics of young individuals.

\begin{figure}[H]
\centering
\includegraphics[scale=.19]{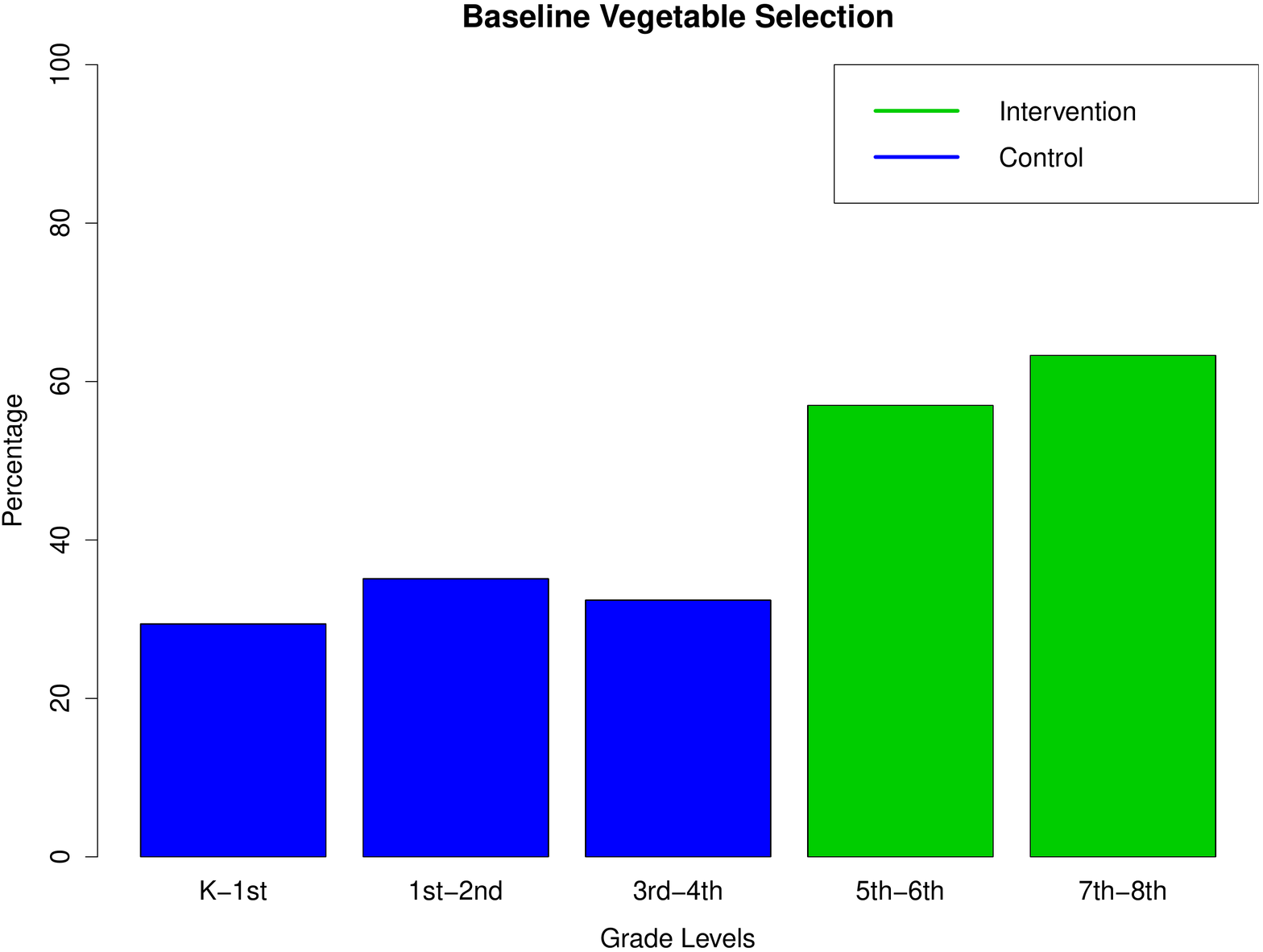}
\hspace{0.5px}
\includegraphics[scale=.19]{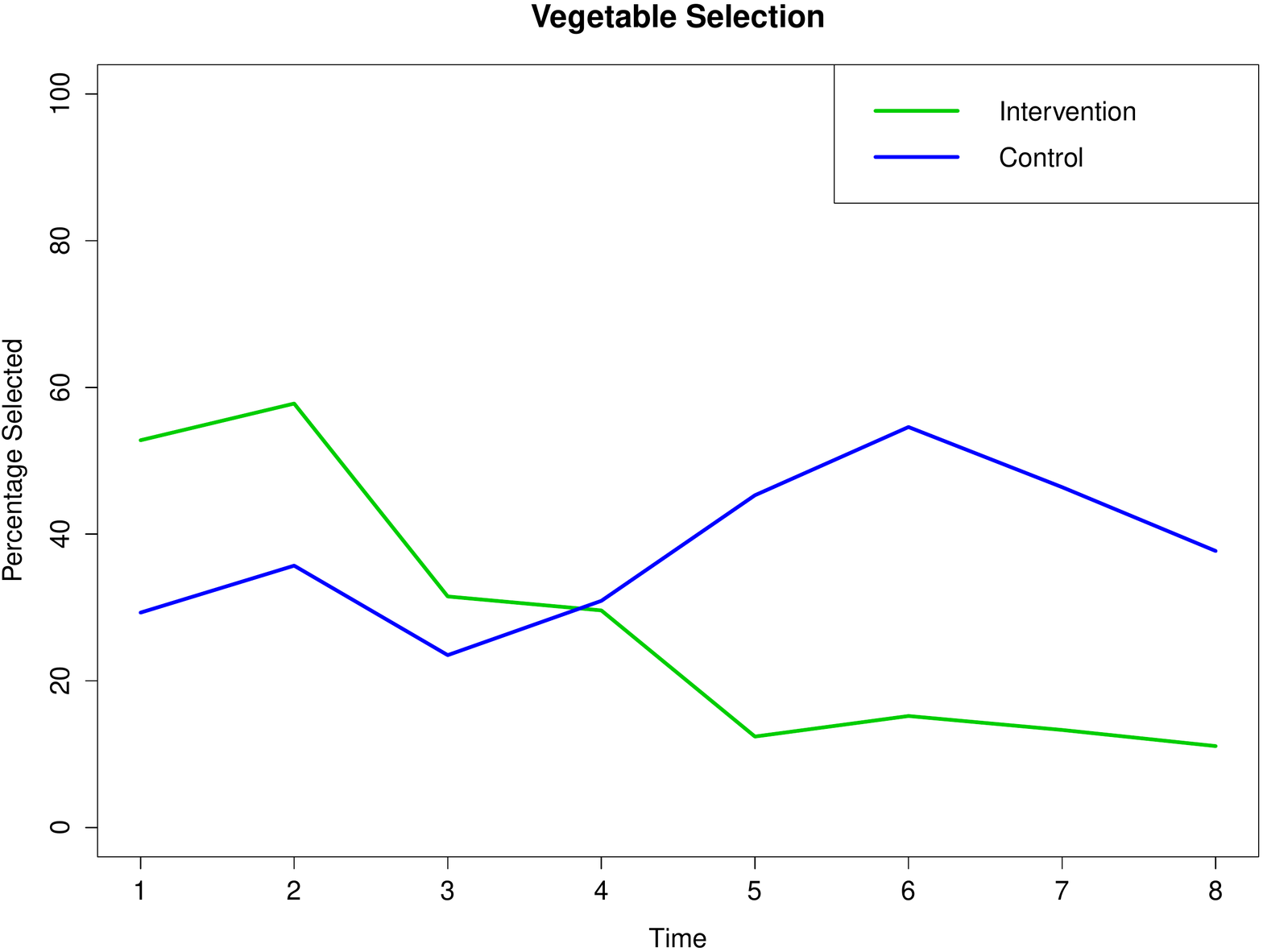} 
\hspace{0.5px}
\includegraphics[scale=.19]{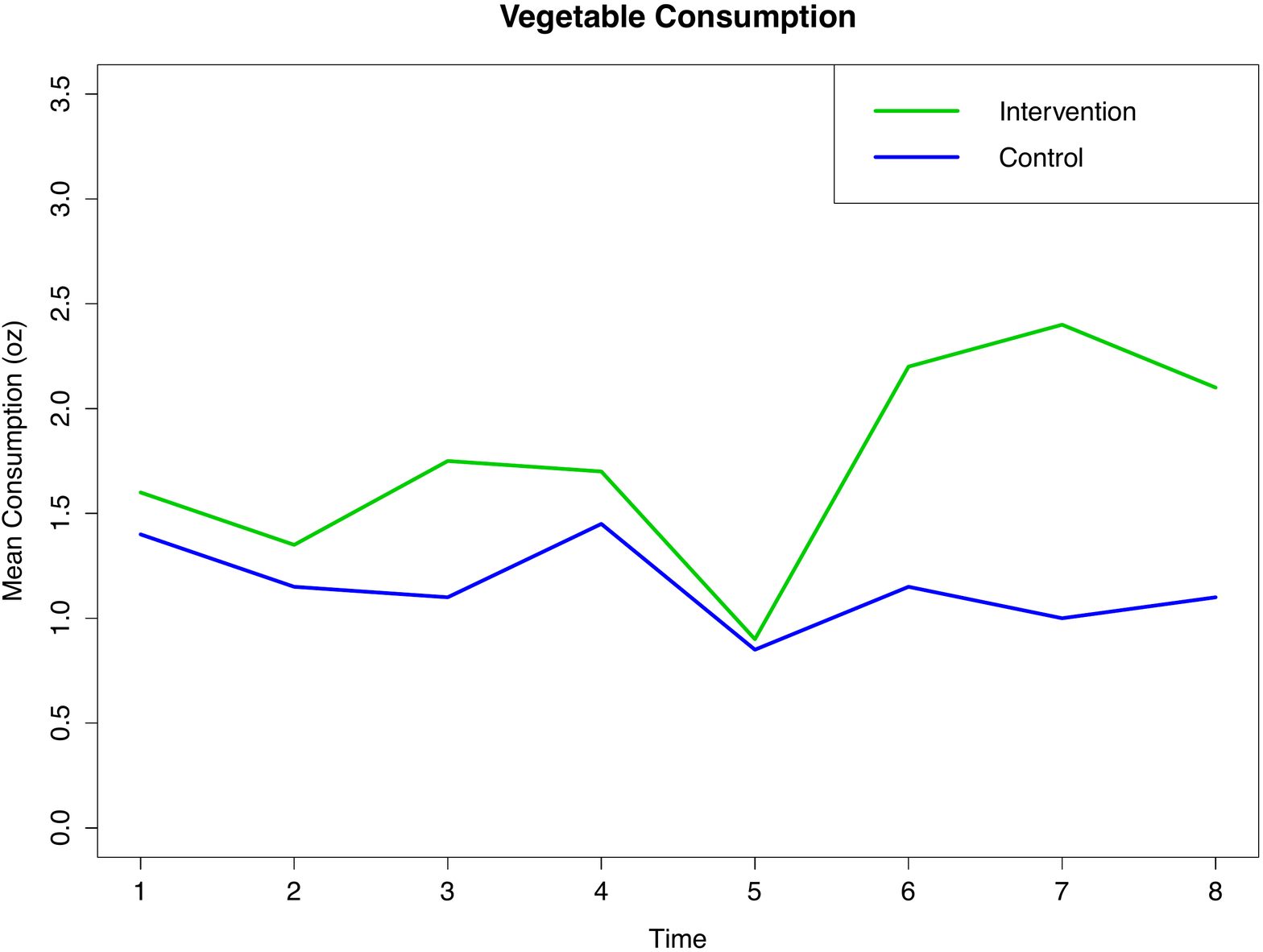} 
\caption{Vegetable Selection and Consumption. Days 1 and 2 represent baseline, days 3 to 6 are the intervention period, and days 7 to 8 represent the testing period.}
\label{fig:anlcdata1}
\end{figure}

Building `positive' childhood memories has been identified as a possibly influential force on the long-term eating behaviors of adults based on the study in \citep{wadhera2015perceived}. Food preferences has been shown to increase with exposure, tasting (not just smelling or seeing), and a positive social experience \citep{birch1980influence,birch1987role}. However, the unpalatability of healthier foods and the onset of neophobia, or the fear of trying something new, influences childrenÕs food choices and can ultimately lower both dietary variety \citep{dovey2008food,falciglia2000food} and the consumption of fruits, vegetables, and meats \citep{cooke2007importance,howard2012toddlers}. These issues have been addressed via exposure techniques (six to ten) \citep{wadhera2015teaching}, where familiarizing children with these foods can improve the liking for and intake of novel foods among preschool and school-aged children \citep{liem2004sweet,sullivan1990pass}. An alternative  approach for increasing liking for and consumption of vegetables is food association learning, in which, a classical conditioning paradigm is applied and considered successful when liking for a novel flavor occurs due to its pleasurable association with the calories or the liked flavor (flavor-flavor learning) it was paired repeatedly with \citep{capaldi1996we}. Although few studies have shown that associative conditioning more effectively increases liking and consumption of vegetables, compared to exposure \citep{capaldi2014associative,fisher2012offering,wadhera2015teaching}; its impact has been minimally studied at the population-level. In our pilot study [\textit{Manuscript in progress}], we studied the effect of associative conditioning among Arizona students. Among the Pre-Kindergarten to 8th grade participants, we found that our method of food association learning acted as a positive reinforcement for children who may be more likely to eat vegetables but did not improve selection or consumption for those who may be more reluctant to eat vegetables (see Figure~\ref{fig:anlcdata1}). These results are utilized as an initial exploration of food association and food preference learning in schools.

The prevalence of childhood (10.4\%) and adult (25.9\%) obesity in Arizona is only slightly lower than national estimates \citep{ADHS(2012a),ogden2014prevalence}; and the 2012 Behavior Risk Factor Surveillance Survey (BRFSS) estimated 60\% overweight or obese adults, 37.5\% of obese adults living in households with food assistance (WIC, SNAP, and/or Free and Reduced Lunch), and increased adult obesity risk among non-daily consumers of fruits and vegetables (30.3\% and 31.7\%) compared to daily consumers (24.6\% and 25.6\%) \citep{ADHS(2012a)}. Although these health disparities are not studied here explicitly, the study of nutrition programs is essential for improving the overall health of Arizona residents. In U.S. children, obesity was higher among Mexican-American (28.8\% boys and 17.4\% girls) and non-Hispanic black (19.8\% boys and 29.2\% girls) than non-Hispanic white (16.7\% boys and 14.5\% girls) \citep{Ogden(2010a)}. Arizona residents comprises demographic characteristics (age, gender, income, education, and employment status) generalizable to the nation \citep{USCensus2013}. However, the presence of food deserts and the economical and environmental barriers puts vulnerable population, or 14.3\% of low-income children, Hispanic (29.9\% in A.Z. and 16.6\% in the U.S.), and American Indian or Alaska Native (4.0\% in A.Z. and 0.7\% in the U.S.) \citep{USCensus2013} residents, at increased risk for insufficient consumption of essential nutrients or overconsumption of unhealthier foods high in saturated and trans fats. 


Though multiple levels of detail and heterogeneity can be incorporated, such an approach could invariably lead to highly complex nonlinear models that would be difficult to analyze. In this first effort, we proceed to study the impact of the three stated factors: dietary programs, social environment, and the physical environment on the distribution of eating patterns. This effort by no means attempts to minimize or underpinned the complexities and challenges associated with understanding the forces behind the dynamics of obesity. What we are trying to do is to introduce a framework for the study of the impact of these three components on the dynamics of obesity under highly simplified conditions at the population-level. We don't expect the results of these caricature models to offer solutions. Our hope is that the population-level framework introduced, its analysis, and the interpretation of the model results would inspire others to expand and improve on this work so that a solid and tested framework would be eventually developed.

\singlespacing

\section{The Mathematical Modeling Framework}
\label{mathmodels}
We develop two models to shed some light on how the interactions among individual factors, the sociocultural environment, and nutrition programs impact the dynamics of eating behaviors and distribution of eaters in school settings. A typical school population can be considered to be composed of two types of students: Ômoderately healthyÕ individuals, denoted $M(t)$, or those who eat a `moderate' amount of fruits, 100\% fruit juice, or vegetables (FJV) (25- 50\% of ``My Plate" guidelines) and the `less' healthyÕ individuals, denoted $L(t)$, or those who eat a `low' amount of FJV (less than 25\% of ``My Plate" guidelines). The first model considers the simplest scenario, where school nutrition programs influence some $L$-eaters to modify their diets to become $M$-eaters but remain in the same environment. However, prior field studies suggest the impact of nutrition education is low and hence this recovery is temporary, suggesting that $M$-eaters can break their `good' diet, a form of recidivism. The second model, incorporates the impact of `positive' food association learning via a \textit{Ratatouille} effect. Both $M$- and $L$-eaters can enter a program, in which, some students learn food association techniques, denoted $A(t)$, where eventually some proportion will develop sustainable food preferences, denoted $P(t)$. In these next subsections, we describe each model, corresponding results, and the conditions under which the diet dynamics are altered.

\subsection{Absence of Food Association, Brief Recovery, and Recidivism}
\label{mathmodel1}
\begin{figure}[H]
\centering
\includegraphics[scale=.35]{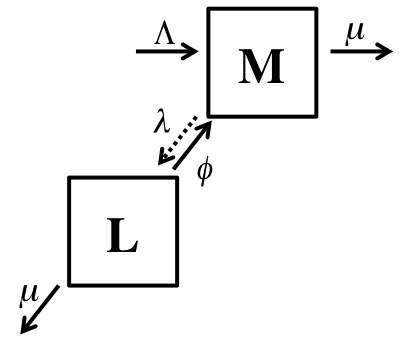}
\caption{A schematic diagram of the healthy eating model 1.}
\label{fig:model1}
\end{figure}

The total population of students, denoted $N$, is made up of $M$- and $L$-eaters. The average time that a student spends in Pre-Kindergarten to 8th grades (10 years) is denoted $1/\mu$. A proportion of $L$-eaters can shift to $M$-eaters after exposure to a nutrition program, denoted $\phi$,  which means that $L$-eaters shift to $M$-eaters but do not change eating environments. The average time that an individual spends in the $L$-eater state before returning to the $M$-eater state is ${1}/{\phi}$. However, the diet changes are temporary due to recidivism since $M$-eaters can shift back to $L$-eaters (see Figure~\ref{fig:model1} for a schematic diagram and Table~\ref{tab:ParamsDefModel1} for variable and parameter definitions). This system is governed by the following equations,
 
 \begin{eqnarray}
M^{\prime} &=& \Lambda - (\lambda + \mu) M + \phi L \nonumber \\
L^{\prime} &=& \lambda M - (\phi + \mu) L, \label{eq:model1}
 \end{eqnarray}
 
\noindent where $\lambda =  \beta {L}/{N}$, represents the fraction of $L$-eaters in the population that interact with $M$-eaters, which in turn, lead to the conversion of $M$- into $L$-eaters at the rate $\beta$, via a social `contagion' process. The contagion process would be considered successful as long as the interactions between $M$ and $L$ lead to an increase in the number of $L$'s. The number of new students entering the school per year is denoted by $\Lambda = \mu N$. 

\singlespacing


\begin{table}[H]
\footnotesize{
\begin{center}
\caption[Model 1 Parameters]{Definition of Model 1 Parameters.}
\begin{tabular}{|llll|}
\hline
\textbf{Parameter} & \textbf{Value} & \textbf{Unit} & \textbf{Description} \\ \hline 
$M$& 0.9 & dimensionless & Proportion of `Moderately' healthy individuals   \\
& & & (consume 25-50\% of ``My Plate" guidelines)    \\
$L$& 0.1 & dimensionless & Proportion of `Less' healthy individuals   \\
&  & &  (consume less than 25\% of ``My Plate" guidelines)    \\
$\beta$& 1.8 & $\frac{1}{year}$ & Peer influence rate shifting a $M$- to an $L$-eater    \\
$\phi$& varies & $\frac{1}{year}$ & Exposure to nutrition programs   \\
$\mu$& 0.10 & $\frac{1}{year}$  & Per-capita student entry and removal rate  \\ \hline
\end{tabular}
\label{tab:ParamsDefModel1}
\end{center}
}
\end{table}

\begin{figure}[H]   
\centering    
    \includegraphics[scale=.35]{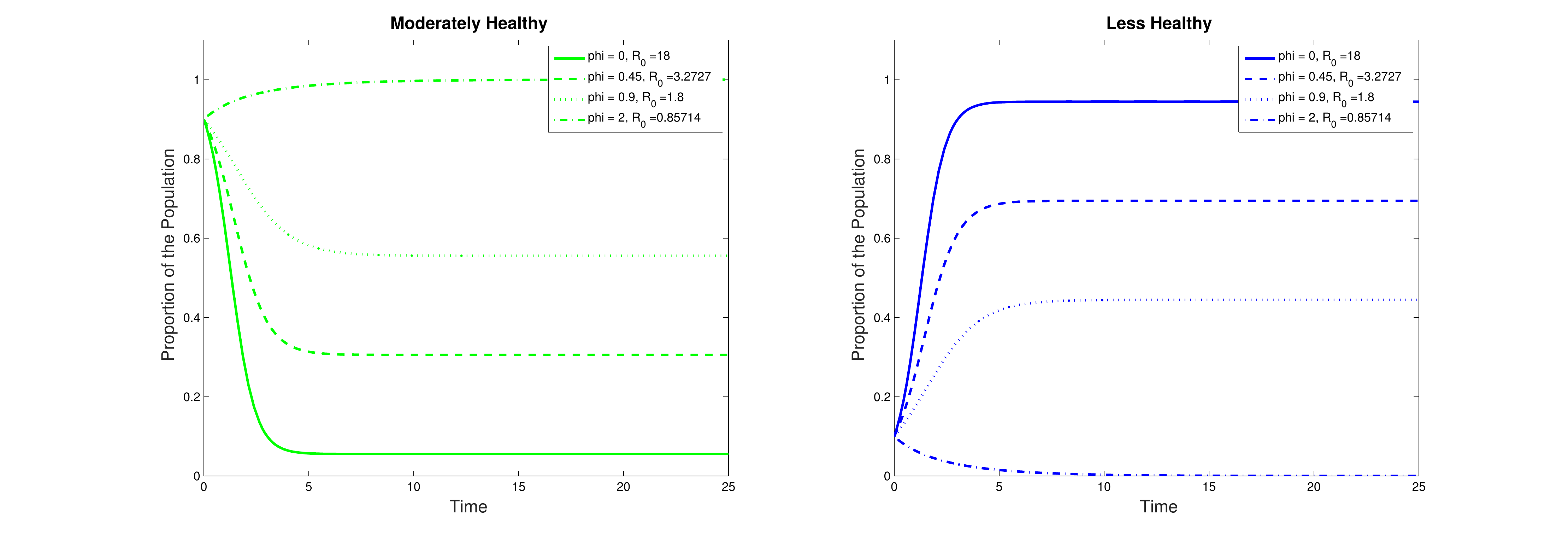}
    \caption{The rate of conversion from $L$- to $M$-eaters ($\phi$) are varied. There is a minimal impact on the proportion of $L$-eaters compared to no education program (solid), where increasing the implementation yields mostly $M$-eaters (dashed dotted). }
     \label{model1plots}
\end{figure}

\noindent The control reproduction number, ${\mathcal{R}_{c,1}}$, is a threshold value permitting the assessment of true success of a nutrition education program. Here, it is defined as a function of the nutrition education program rate $\phi$, 
\[{\mathcal{R}_{c,1}(\phi)} = \frac{\beta}{\mu + \phi}, \]
\noindent where ${1}/{(\mu + \phi)}$ represents the total average time spent in the district as an $L$-eater before shifting to an $M$-eater following a nutrition education program. When there is no nutrition education program, that is $\phi=0$, then ${\mathcal{R}_{c,1}(\phi)}$ becomes, ${\mathcal{R}_{c,1}(0)} = \frac{\beta}{\mu},$
that is, the threshold becomes the product of $\beta$, the effective conversion rate per $L$, and $1/\mu$, the average time a student remains in the education system. The above simplistic model will not be used to highlight the effectiveness or lack thereof of nutrition education on altering the prevalence of $L$-eaters. However, this model assumes that the educational effort (per person) modelled by $\phi$ remains part of the culture and it is continuously implemented. Our pilot data [\textit{Manuscript in Progress}] suggested that $L/N = 0.7$ (i.e., 70\%), hence at equilibrium $L/N = 1 - 1/{\mathcal{R}_{c,1}(0)}$ and $M/N = 1/{\mathcal{R}_{c,1}(0)}$. With $1/{\mathcal{R}_{c,1}(0)} = 0.3$ and $1/\mu = 10$ years, we can estimate $\beta/\mu = 1/0.3$, or $\beta = (1/0.3) \cdot (1/10) = 1/3 \simeq 0.33$. Using these values, model simulations show that increasing the nutrition programs, $\phi$, will decrease the proportion of ${L}/{N}$ eaters (see Figure~\ref{model1plots}). A sociocultural environment with mostly $M$-eaters is achieved for large values of $\phi$. If ${\mathcal{R}_{c,1}(\phi)} > 1$, then the amount of $L$-eaters would increase with the proportion of non-converts decreasing. In the long-term, the model would achieve a steady state, that is, the student population will settle into a `fixed' proportion of $L$-eaters ($L/N$) and $M$-eaters ($M/N$). If ${\mathcal{R}_{c,1}(\phi)} < 1$, then the population would consist of mostly $M$-eaters instead of $L$-eaters in the long-run. The system is rescaled such that $X = M/N$, $Y=L/N$, and $N/N = X + Y = 1$. There are two equilibrium points (in proportions) are:  the diet-problem-free state 
\[E_{0,1} = (X_{0,1}, Y_{0,1})^{\prime} = (1,0)^{\prime}\]
\noindent  and the diet-problem-endemic state 
\[E_{1,1} =(X_{1,1}, Y_{1,1})^{\prime} = \left( \frac{1}{\mathcal{R}_{c,1}} , 1- \frac{1}{\mathcal{R}_{c,1}} \right)^{\prime}.\]
\noindent The prime ${}^{\prime}$ here denotes vector transpose. We claim that $E_{1,0}$ is globally asymptotically stable \textit{if and only if} $\mathcal{R}_{c,1} \leq 1$ while $E_{1,1} $ is globally asymptotically stable whenever it exists (i.e., \textit{if and only if} $\mathcal{R}_{c,1} > 1$). 
Hence, the inequality $\mathcal{R}_{c,1} \leq 1$ is equivalent to 
\begin{equation*}
\frac{1}{\phi} \le \frac{1/{\mu}}{\mathcal{R}_{c,1}(0) - 1}.
\end{equation*}
This means that the shorter the average time spent in the $L$-eater state is, the better chance we have to eliminate the diet problem at the population-level.

\singlespacing

\subsection{Ratatouille Effect}
\label{mathmodel4}
\begin{figure}
\centering
\includegraphics[scale=.35]{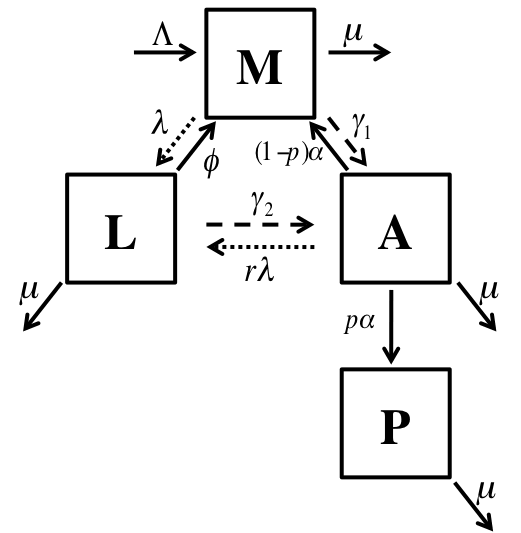}
\caption{A schematic diagram of the eating behavior model 2.}
\label{fig:model4}
\end{figure}

A slightly modified version of Model (\ref{eq:model1}) permits the study of food association learning with varying levels of effectiveness. Here, $M$-eaters will enter the food association learning program at the per-capita rate $\gamma_1$. After association learning, a portion $p$ will become food preference learners ($P$-eaters) at the combined rate $p \alpha $, in which, we consider the food association learning program successful. Recidivism of $A$-eaters, where they return to old ways of eating as $M$-eaters, occurs at the rate $(1- p)\alpha $ or as $L$-eaters after social interactions with $L$-eating peers at rate  $r\lambda$, where $\lambda = \beta {L}/{N}$. The $M$-eaters who do not enter the food association program would either maintain current eating habits or by social interactions with peers, $\lambda$, they would become $L$-eaters. Finally, $L$-eaters can shift to $M$-eaters at rate $\phi $ or they join the food association program and therefore transit to $A$-eaters at rate $\gamma_2$ (see Table~\ref{tab:ParamsDefModel4} for variable and parameter definitions and Figure~\ref{fig:model4} for a schematic diagram). This new model is governed by the following equations,
\begin{eqnarray}
M^{\prime} &=& \Lambda - (\mu + \lambda + \gamma_1) M + \phi  L + (1-p)\alpha  A, \nonumber  \\
A^{\prime} &=& \gamma_1 M + \gamma_2 L- (r\lambda + \mu + \alpha )A,  \label{eq:model4}  \\
L^{\prime} &=& \lambda M - (\phi  + \gamma_2 + \mu)L + r\lambda A, \nonumber\\
P^{\prime} &=& p\alpha  A - \mu P \nonumber
\end{eqnarray}

\noindent where the total population is $N = M + L + A + P$ and student school entry rate is $\Lambda = \mu N$ 

\singlespacing

\begin{table}
\footnotesize{
\begin{center}
\caption[Model 2 Parameters]{Definition of Model 2 Parameters.}
\begin{tabular}{|llll|}
\hline
\textbf{Parameter} & \textbf{Value} & \textbf{Unit} & \textbf{Description} \\ \hline 
$M$& 0.4 & dimensionless & Proportion of `Moderately' healthy individuals   \\
& & & (consume 25-50\% of ``My Plate" guidelines)    \\
$L$& 0.1 & dimensionless & Proportion of `Less' healthy individuals   \\
&  & &  (consume less than 25\% of ``My Plate" guidelines)    \\
$A$& 0.2 & dimensionless & Proportion of food association learners    \\
$P$& 0.3 & dimensionless & Proportion of food preference learners   \\
$\beta$& 1.8 & $\frac{1}{year}$ & Peer influence rate shifting a $M$- to an $L$-eater    \\
$\gamma_1$& 0.35 & $\frac{1}{year}$ & Entry rate into food association program as an $M$-eater  \\
$\gamma_2$& 0.06 & $\frac{1}{year}$ & Entry rate into food association program as an $L$-eater  \\
$p$& varies & dimensionless & Proportion those who become "preference learners"   \\
$\alpha $ & 0.4 & $\frac{1}{year}$ & Effectiveness rate of food association learning \\
$\phi $ & 0.6 & $\frac{1}{year}$ & Recidivism rate from a $L$- to an $M$-eater \\
$r$ & 0.1 & dimensionless & Denotes the relative susceptibility of $A$-eaters with respect \\
& & &to $M$-eaters who shift to an $L$-eater \\
$\mu$& 0.10 & $\frac{1}{year}$  & Per-capita student entry and removal rate  \\ \hline
\end{tabular}
\label{tab:ParamsDefModel4}
\end{center}
}
\end{table}

\noindent Model (\ref{eq:model4}) is rescaled in terms of sub-population proportions: $X = M/N, W = A/N, Y = L/N$, and $Z=P/N$. The diet-problem-free equilibrium is $E_{0,2} = (X_{0,2}, W_{0,2}, 0, Z_{0,2})^{\prime}$, where
\begin{eqnarray*}
X_{0,2} & = & \frac{\mu(\alpha  + \mu)}{(\alpha  + \mu)(\gamma_1 + \mu) - (1-p)\alpha  \gamma_1},\\
W_{0,2} & = & \frac{\mu \gamma_1}{(\alpha  + \mu)(\gamma_1 + \mu) - (1-p)\alpha  \gamma_1},\\
Z_{0,2} & = & 1 - X_{0,4} - W_{0,4}.
\end{eqnarray*}

\begin{figure}
\centering    
    \includegraphics[scale=.35]{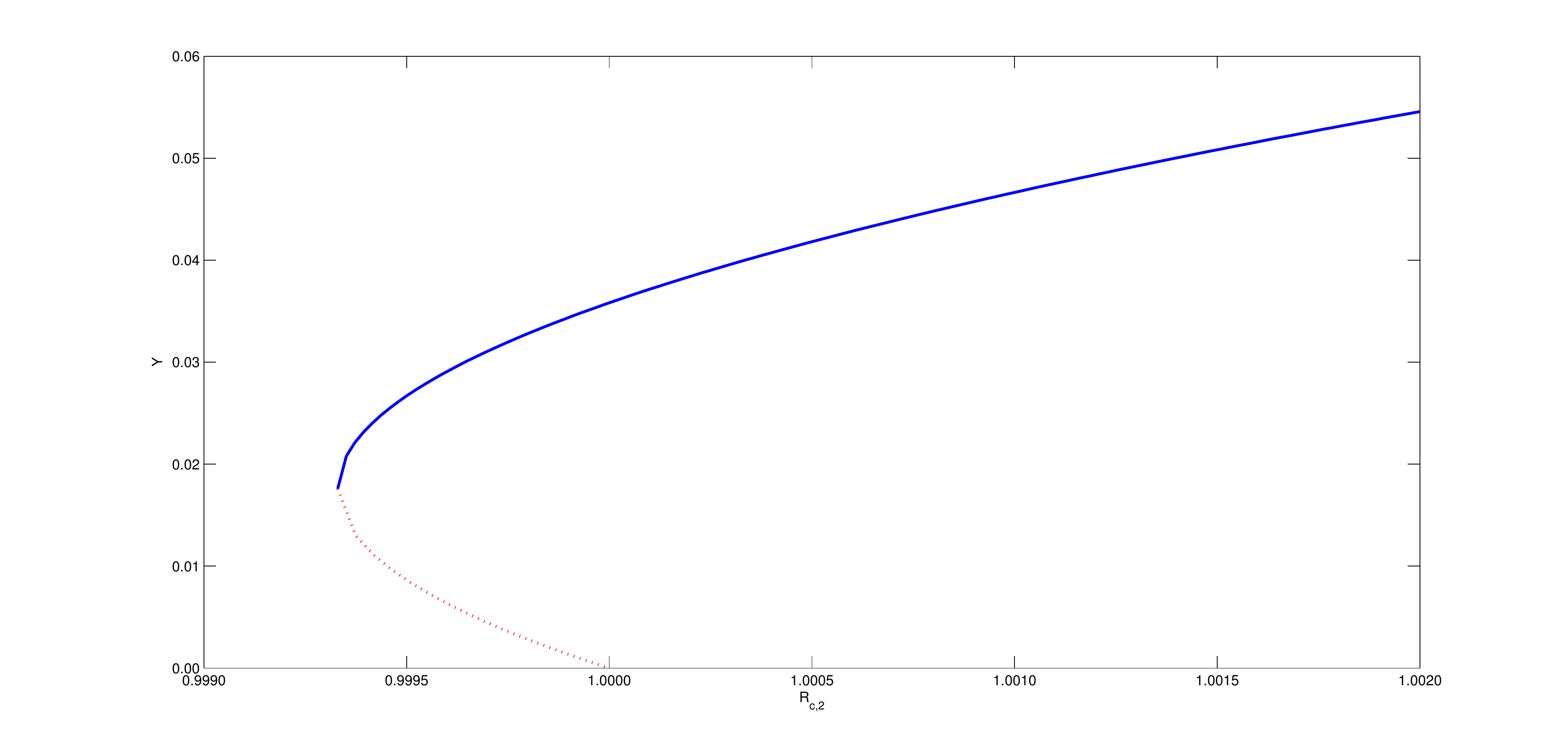}
    \caption{The proportion of unhealthy-eaters, or $L$-eaters, at equilibrium as a function of the control reproduction number $\mathcal{R}_{c,2}$. Simulations are done for $\mu = 0.1, \gamma_1 = 0.35, \gamma_2 = 0.06, \alpha  = 0.4, \phi  = 0.06, r = 0.8258$ and $p = 0.9298$.}
     \label{fig:RCy5}
\end{figure}

\noindent  It is locally asymptotically stable \textit{if and only if}  $\mathcal{R}_{c,2} < 1$, where $\mathcal{R}_{c,2} = (1-q)\mathcal{R}_{c,1}$
is the control reproduction number for the model with the \textit{ratatouille} effect. The proportion, $q = \frac{(p\alpha + (1-r)\mu)\gamma_1}{\mu(\alpha + \mu + \gamma_1) + p\alpha \gamma_1}$,
represents the reduction in the control reproduction number $\mathcal{R}_{c,1}$ due to the application of the education association program. The analysis reveals further that the rescaled model shows the existence of subcritical endemic states (backward bifurcation phenomenon) \textit{if and only if} the following set of inequalities is held

\begin{equation}
\phi  > \phi^c, \quad r_1 < r < r_2, \quad p > p^c , \label{Model4BB}
\end{equation}
that is, if $\frac{1}{\phi}$ is small enough, the susceptibility is within some pre-specified range, and the proportion of preference learners is high enough 
where
\begin{eqnarray*}
\phi^c & = & \frac{\mu(\gamma_2 + \gamma_1 + 2 \mu) + 2(\gamma_1 + \mu)\sqrt{\mu(\gamma_2 + \mu)}}{\gamma_1},\\
r_1 & = & \frac{\alpha  + \mu}{2(\phi  + \mu + \gamma_1)} \left[\frac{\gamma_2}{\gamma_1} + \frac{\phi }{\mu} - 1 - \sqrt{{\left( \frac{\gamma_2}{\gamma_1} + \frac{\phi }{\mu} - 1 \right)}^2 - 4{\left(\frac{\gamma_2 + \mu}{\gamma_1}\right)} \left(1 + \frac{\gamma_1 + \phi }{\mu} \right)}\right],\\
r_2 & = & \frac{\alpha  + \mu}{2(\phi  + \mu + \gamma_1)} \left[\frac{\gamma_2}{\gamma_1} + \frac{\phi }{\mu} - 1 + \sqrt{{\left( \frac{\gamma_2}{\gamma_1} + \frac{\phi }{\mu} - 1 \right)}^2 - 4{\left(\frac{\gamma_2 + \mu}{\gamma_1}\right)} \left(1 + \frac{\gamma_1 + \phi }{\mu} \right)}\right],\\
p^c & = & \frac{\mu\left[(r\gamma_1 + \alpha  + \mu)^2 + (r - 1)[r\gamma_1(\phi +\mu) - \gamma_2(\alpha +\mu)] \right]}{\alpha [r\gamma_1(\phi +\mu) - \gamma_2(\alpha  + \mu)]}.
\end{eqnarray*}
Thus, if Condition (\ref{Model4BB}) holds, then the model has two diet-problem-endemic equilibria for $\mathcal{R}_{c,2} < 1$. Figure \ref{fig:RCy5} shows the bifurcation diagram for the ratatouille model in the plane $(\mathcal{R}_{c,2}, Y)$, where the solid curve corresponds to a diet-problem-endemic equilibrium with higher level of the endemic prevalence of $L$-eaters and the dotted curve corresponds to a diet-problem-endemic equilibrium with lower level of $L$-eaters' endemic prevalence, and both exist when $\mathcal{R}_{c,2} < 1$. Further, as $\mathcal{R}_{c,2}$ decreases, both curves approach each other until reaching the turning point \cite{safan2006minimum} at which both of them coalesce. The value of the control reproduction number at this turning point is given by $\mathcal{R}_{c,2}^{\star 1}$, where
\begin{equation}
\mathcal{R}_{c,2}^{\star 1} = \frac{(\alpha  + \mu + r\gamma_1)[\gamma_2(p\alpha  + \mu) +\mu (r(\phi  + \mu - \gamma_1) - (\alpha  + \mu)) + 2\sqrt{D_{R_{c,2}}}]}{r(\phi  + \gamma_2 +\mu) [(\alpha  + \mu)(\gamma_1 + \mu) - (1-p)\alpha \gamma_1]}
\end{equation}
and
\begin{equation*}
D_{R_{c,2}} = r\gamma_1\gamma_2\mu^2 + \mu[p\alpha  + (1-r)\mu][r\gamma_1(\phi  + \mu) - \gamma_2(\alpha  + \mu)].
\end{equation*}

\singlespacing

\noindent In fact, the value $\mathcal{R}_{c,2} = \mathcal{R}_{c,2}^{\star 1}$ is, a threshold value, that determines the nonexistence and existence of diet-problem-endemic states. If at least one of the conditions (\ref{Model4BB}) is not satisfied, then the model shows the existence of forward bifurcation (supercritical endemic state), in which, a unique diet-problem-endemic equilibrium exists and is stable for $\mathcal{R}_{c,2} > 1$, while no endemic state exists for $\mathcal{R}_{c,2} < 1$. Hence,  $\mathcal{R}_{c,2} = 1$ is the threshold level that indicate the nonexistence and existence of diet-problem-endemic states. Thus, we summarize the above results as follows: the critical control reproduction number below which diet-problem-endemic equilibria do not exist is given by
\begin{equation}
\mathcal{R}_{c,2}^{\star} = \begin{cases}
\mathcal{R}_{c,2}^{\star 1} \qquad& \text{if the bifurcation is backward},\\
1 \qquad& \text{if the bifurcation is forward}.
\end{cases}\label{critbeta}
\end{equation}

\subsubsection*{\protect\normalsize Diet-problem containment possibility}
\begin{figure}[H]   
\centering    
    \includegraphics[scale=.35]{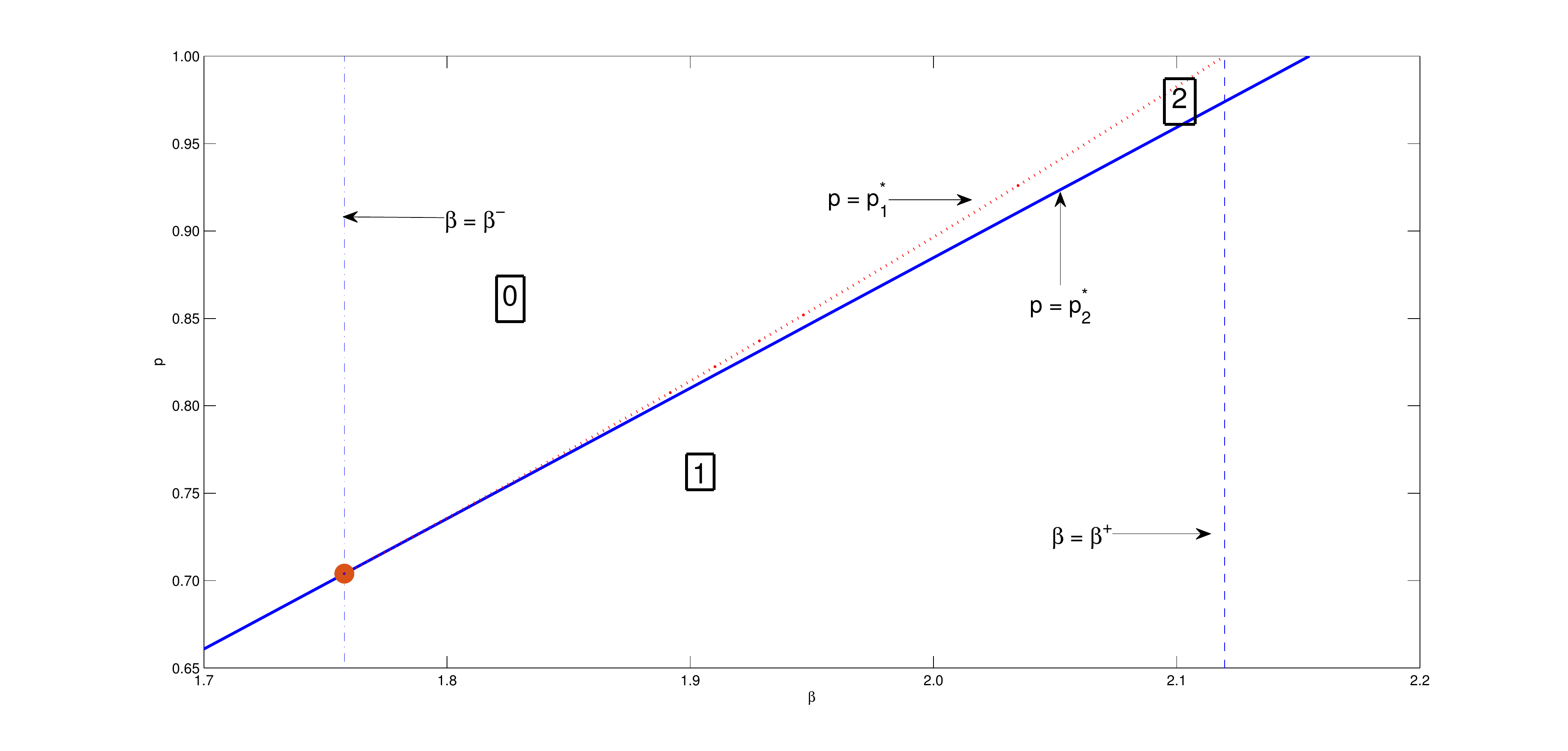}
    \caption{The critical probability of effectiveness $p^{\star}$ subdivides the $(\beta, p)$ plane into regions (denoted by $0, 1,$ and $2$) according to the number of diet-problem-endemic equilibria. }
     \label{fig:betapplane}
\end{figure}

Addressing the possibility of containing (getting rid of) the diet problem is certainly of utmost importance. Here, we discuss the existence of necessary and sufficient conditions required to eliminate the diet-endemic problem based on the implementation of a food association program with effectiveness $p \in [0, 1]$. In the literature of mathematical epidemiology, the basic reproduction number $\mathcal{R}_0$ is a key concept, the public health cornerstone used to determine the minimum effort required to eliminate an infection when the model doesn't exhibit the existence of multiple endemic equilibria. 
However, in the last two decades several models exhibited bistable endemic states, where backward bifurcation and hysteresis phenomena are shown to exist. In such cases, $\mathcal{R}_0 < 1$ is a necessary but not sufficient condition for eliminating the infection. For a model with backward bifurcation, it has been shown in \cite{safan2006minimum} that the ratio $\mathcal{R}_0/\mathcal{R}_0^{\star}$ could be interpreted as a reproduction number and so, reducing this ratio to below one ensures an effective control of the infection. Thus, if we solve the inequality  $\mathcal{R}_{c,2}/\mathcal{R}_{c,2} ^{\star} < 1$ in terms of the probability $p$, we get

\begin{equation}
p > p^{\star} = \begin{cases}
p_1^{\star} \qquad& \text{if the bifurcation is backward},\\
p_2^{\star} \qquad& \text{if the bifurcation is forward}
\end{cases}\label{critp}
\end{equation}
where 
\[p_1^{\star}  =  1 - \frac{1}{\alpha  \gamma_2^2}\left[ Q_1 + \sqrt{Q_1^2 - \gamma_2^2 Q_2}\right], \indent p_2^{\star} = 1 - \frac{1}{\alpha \gamma_1} \left[ (\alpha  + \mu)(\gamma_1 +\mu) - \frac{\mu \beta(\alpha  + \mu + r\gamma_1)}{\phi  + \mu}\right] \]
and 
\begin{eqnarray*}
Q_1 & = & \gamma_2 [(\alpha  + \mu)(\gamma_2 +\mu) + r\mu(\phi  + \gamma_1 + \mu - \beta)] - 2r\mu\gamma_1(\phi  + \gamma_2 + \mu),\\
Q_2 & = & [(\alpha  + \mu)(\gamma_2 +\mu) + r\mu(\phi  + \gamma_1 + \mu - \beta)]^2 - \\
        &  &  4r\mu[(\phi  + \gamma_2 +\mu)(\alpha  + \mu)(\gamma_1 + \mu) - \mu\beta(\alpha  + \mu + r\gamma_1)].
\end{eqnarray*}
\noindent Formula (\ref{critp}) determines the critical probability ($p^{\star}$) of effectiveness of a food association program above which the diet-problem-endemic state(s) disappear. Figure~\ref{fig:betapplane} shows the critical level of the food association effectiveness $p^{\star}$ as a function of the contact rate $\beta$. The vertical line $\beta = \beta^-$ separates between nonexistence and existence of a backward bifurcation. Therefore, for $\beta \le \beta^-$, the curve $p = p_2^{\star}$ separates between existence and nonexistence of diet-problem-endemic equilibria. Thus, a probability of effectiveness slightly above $p_2^{\star}$ ensures an effective control of the diet-endemic problem. However, if $\beta^- < \beta < \beta^+$, then backward bifurcation exists and $p = p_1^{\star}$ is the threshold above which diet-problem-endemic equilibria do not exist. Thus, a food association program with probability of effectiveness slightly higher than $p_1^{\star}$ exhibits a die-out of the diet-endemic problem, where
\begin{eqnarray*}
\beta^- &=& \phi  +\mu - \gamma_1- \frac{\alpha  + \mu}{r} + \frac{\gamma_2}{r}\left( 1 + \frac{\alpha }{\mu} p^c \right) ,\\
\beta^+ & = & \phi  +\mu - \gamma_1 - \frac{\alpha  + \mu}{r}\left(1 - \frac{\gamma_2}{\mu}\right) + 2\sqrt{\frac{\gamma_1\gamma_2}{r} + \frac{\alpha  + (1-r)\mu}{r\mu} \left(\gamma_1(\phi  + \mu) - \frac{\gamma_2(\alpha  + \mu)}{r} \right)}.
\end{eqnarray*}

\noindent Here, the level $\beta = \beta^+$ represents the value at which $p_1^{\star}$ hits the upper bound $p=1$. Thus, for $\beta > \beta^+$, there is no feasible value of $p$ that ensures a wash out of the diet-endemic problem and we may seek another control strategy to first reduce the contact rate $\beta$ to below $\beta^+$ and then apply a food association program with high enough probability of effectiveness. This ensures an effective control of the diet-problem.

\singlespacing

\begin{figure} [H]   
\centering    
    \includegraphics[scale=.35]{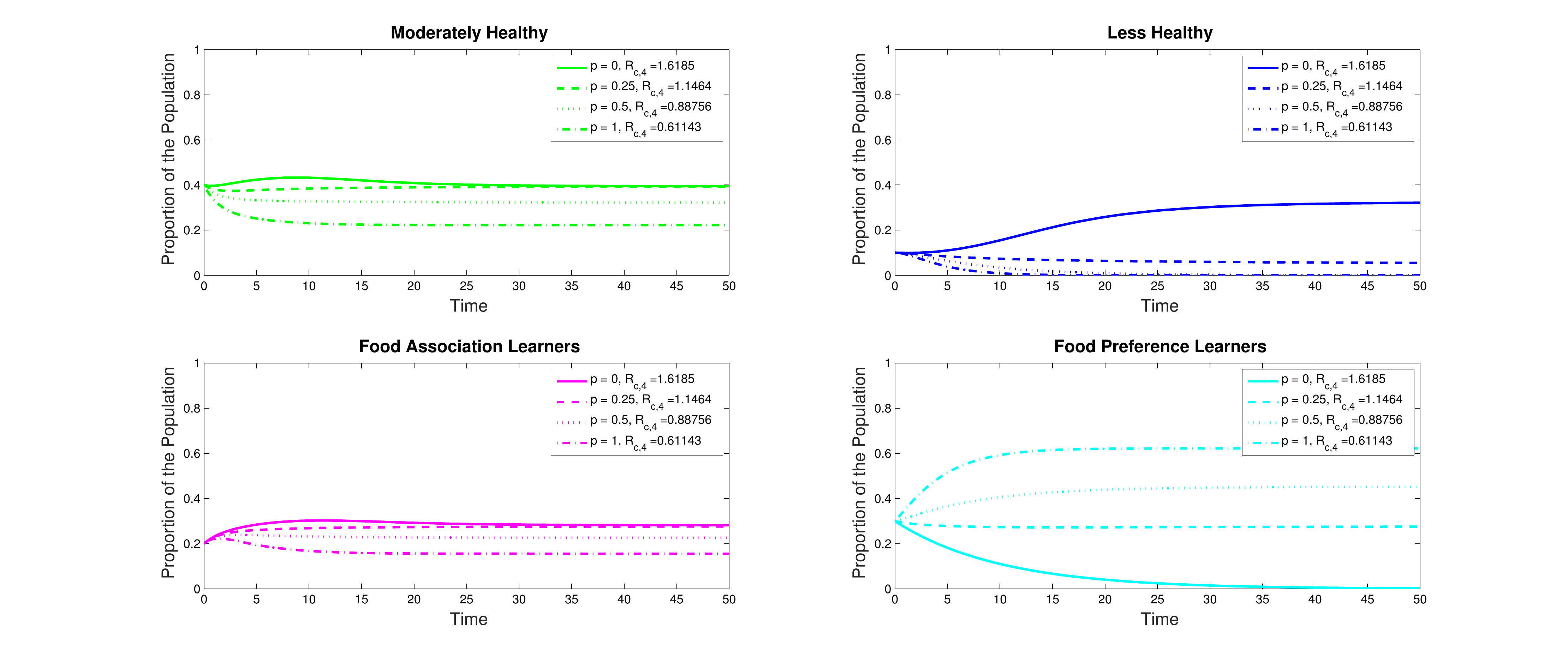}
    \caption{Time series analysis for the subpopulation proportions for different values of the food association efficacy probability $p$ and the control reproduction number $\mathcal{R}_{c,2}$.}
     \label{fig:model4timeseries}
\end{figure}

\noindent  Figure~\ref{fig:model4timeseries} shows a time series analysis for the model for a fixed $\beta$ and four different levels of $p$. The proportion of $L$-eaters approaches zero when $p=0.5$ and $p =1$, while when $p=0$ and $p=0.25$, it approaches a constant value. This implies that if the efficacy of the program is 50\% or greater, then the $M$- and $A$-eaters are reduced, while $L$-eaters approach zero, and $P$-eaters are largest, compared to a program with lower efficacy ($p<0.5$). Hence, a food association program that leads to food preference learning can be an effective nutrition intervention strategy. However, this would require knowledge on the culture, norms, and values of the community to create and implement such a program. 

\singlespacing

%


\section{Discussion}
\label{discussion}
The goals of many nutrition programs are to instill healthy and sustainable eating habits among young individuals. Since food association learning has been identified as a more effective approach, we study its potential impact through use of mathematical models. Two models were developed in order to study eating behavior learning and the resulting diet dynamics of young individuals. The first model considered the case when there is no food association learning program and the second incorporated food association and food preference learning. Results of Model 1 indicate that some nutrition program at schools are better than none at all. If effective, or $p$ large enough, then the food association learning program is a potential impactful strategy at reducing the proportion of $L$-eaters shown by the results of Model 2. These results demonstrates the importance of nutrition education curriculum, learning, and socialization in schools. However, more work is needed to understand how to create and implement an effective program so that it incorporates the culture, norms, and values of the community, supporting the conclusions of other studies  \citep{ADA(1999),perez2001school,perez2003nutrition,story2008creating}. Future work would more effectively incorporate data from the literature. The parameter values we chose (see Tables~\ref{tab:ParamsDefModel1} and \ref{tab:ParamsDefModel4}) were qualitatively estimated based on observations from this pilot study and the literature but more work is needed to quantify these values.

\small
\bibliography{eatingbehaviors}

\end{document}